# Planetary Genealogy

Christoph Burkhardt

University of Münster Institut für Planetologie Wilhelm-Klemm-Straße 10 D-48149 Münster, Germany
burkhardt@uni-muenster.de

## ABSTRACT

The detection of exoplanets and accretion disks around newborn stars has spawned new ideas and models of how our Solar System formed and evolved. Meteorites as probes of geologic deep time can provide ground truth to these models. In particular, stable isotope anomalies in meteorites have recently emerged as key tracers of material flow in the early Solar System, allowing cosmochemists to establish a "planetary isotopic genealogy". Although not complete, this concept substantially advanced our understanding of Solar System evolution, from the collapse of the Sun's parental molecular cloud to the accretion of the planets.

## HOW TO FORM A SOLAR SYSTEM

Until 25 years ago, imagining the shape and makeup of planetary systems around other stars was mostly left to the fantasy of science fiction writers. The few scientists dealing with planet formation at that time mostly devised their theories looking at the one system they knew of – our own. Its architecture with terrestrial planets in the inner part, followed by gas and ice giants outside, was taken as a generic outcome of planet formation among Sun-like stars. This was a pretty biased view. Today, with >4,800 known exoplanets and >800 planetary systems (see exoplanet.eu for the latest numbers), it is evident that the formation and evolution of planetary systems are more diverse than we have ever imagined. For instance, many exoplanets orbit very close to their host stars, and closely spaced planets can have very different properties. These unexpected observations led, for example, to the realization that there are various ways for planets to migrate in the disk of gas and dust in which they form.

Accretion disks around young stars can now be observed in situ with astonishing detail through telescopes such as the *Hubble Space Telescope* or the *Atacama Large Millimeter/submillimeter Array* (*ALMA*). The iconic *ALMA* picture of the young stellar object HL Tauri (see title figure) shows the disk around a star <1 My old, at submillimeter wavelengths. The dark concentric rings in the disk reveal regions of reduced dust density, likely having been carved by the formation of planets. The presence of planets in such young disks was unexpected and challenges preconceived ideas of planet formation and disk evolution.

In response to these novel observations, astronomers have expanded their disk and planet formation models with new mechanisms of accretion and planet migration (Ida 2019). For example, it was realized that so-called "streaming instability" allows dust grains and pebbles to concentrate into dense filaments in the rotating gaseous disk, providing a mechanism for the fast accretion of ~100 km-sized bodies called planetesimals. Likewise, a mechanism known as "pebble accretion" may allow for the formation of planets within the few-million-year lifetime of the gaseous disk by the efficient accretion of sunward-drifting pebbles from outer disk regions. Although initially devised to explain the necessary rapid accretion of the solid cores of gas giants during the lifetime of the disk, pebble accretion is now also discussed as a mechanism for the fast formation of terrestrial planets. This is in stark contrast to the classical model, where terrestrial planets are produced in stochastic collisions between Moon-size to Mars-size planetary embryos over a timescale of ~100 My, long after the gas disk is gone.

Taken together, the new models can account for much of the zoo of observed exoplanetary systems and disk structures. However, this inclusiveness comes at the price of a widely opened-up parameter space. Thus, although the new observations spawned new ideas about how our own Solar System might have formed and evolved, retracing its *actual* evolutionary path by modeling alone is impossible without a time machine. This is where meteorites and cosmochemistry come into play.

## METEORITES AS TIME CAPSULES FROM THE EARLY SOLAR SYSTEM

The next best thing to a time machine, which would allow the Solar System's formation to be observed directly, is having samples from the event. These samples are meteorites. They are debris of collisions of planetary objects and mostly derive from the asteroid belt located between Mars and Jupiter, where a large variety of planetesimals have survived since their formation in the early Solar System. From a petrological perspective, meteorites come in two main 'flavors', representing the nature of their parent body: differentiated, or undifferentiated (FIG. 1).

Differentiated meteorites crystallized from melts and record the conditions and timing of crust, mantle, and core formation in their respective parent bodies. By contrast, undifferentiated meteorites (the chondrites) were never heated to high enough temperatures to induce melting, and thus still contain the primitive materials that were present in the solar nebula during the formation of their parent-bodies. These materials range from high-temperature condensates, such as the Ca,Al-rich inclusions (CAIs) whose age of 4.567 Ga marks the birth of the Solar System on an absolute timescale, to (sub-)millimeter-size igneous spherules (chondrules), formed by transient heating events, all set in a fine-grained matrix that includes primitive organic molecules from the interstellar medium and presolar grains from previous star generations. The relative amounts and characteristics of these materials vary widely between different chondrite groups, resulting in bodies with, for example, different bulk redox states or volatile element contents. Aside from these differences, however, all chondrite groups have relative abundances of nonvolatile elements that are identical, within uncertainty, to those of the Sun's photosphere. Because the Sun contains 99.86% of the mass of the Solar System, chondrites, therefore, represent a proxy for the chemical composition of the solar nebula from which the Sun and planets formed. As such, chondrites also provide the basis for any estimate of Earth's bulk composition, a prerequisite for reconstructing its evolutionary history by geochemical means.

Meteorites contribute to a better understanding of the history of the Solar System and planets in many ways, but studying their isotopic composition is particularly useful.

Radiogenic isotope systems provide absolute or relative age constraints on processes in the solar accretion disk or on planetary bodies, and mass-dependent isotope fractionation can be used to investigate the conditions under which these processes took place. Herein, I will concentrate on isotope variations that are of nucleosynthetic origin. They see through any mass-dependent isotope fractionation, and so make for excellent genetic tracers of planetary materials, from the molecular cloud stage to the formation of planets.

## PLANETARY ISOTOPE GENETICS

Until ~50 years ago, it was a firm assumption that the Solar System started from a well-mixed cloud and that all Solar System materials had the same initial isotopic composition. However, in some chondrite materials isotope variations were detected that could not be explained by radioactive decay, mass-dependent fractionation, or by exposure to galactic cosmic rays during the meteoroids' travel through space (Clayton et al. 1973). Instead, these mass-independent isotope variations appeared to be a diluted version of the isotope effects calculated in models of nucleosynthesis—the process of the formation of the elements. This provided the first hint that meteorites contain isotopically anomalous materials, older than the Solar System itself. After years of detective work, the materials carrying these anomalies were identified as presolar grains (Lewis et al. 1987). These nanometer- to micrometer-size grains condensed in the outflows of dying stars, were part of the Solar System's parental molecular cloud, and have survived in the matrix of chondrites, where they can now be identified by their vastly different isotopic compositions relative to the average Solar System (Nguyen and Messenger 2011).

With improvements in mass-spectrometer precision, it became evident in the last decades that it is not only presolar grains that can exhibit nucleosynthetic isotope anomalies. In highly diluted form such anomalies also appear in other chondrite components and even bulk meteorites (Dauphas and Schauble 2016). For example, CAIs contain slight excesses in nuclides produced by the *r*-process (rapid-neutron capture), a nucleosynthetic pathway associated with the high-neutron density environments of supernovae (the violent end of massive stars running out of fuel) or of kilonovae (mergers of neutron stars or neutron stars and black holes). Although the anomalies in CAIs are orders of magnitude smaller than those in presolar grains, they nevertheless result in resolvable anomalies between chondrite groups with variable CAI abundances. Likewise, some elements in bulk meteorites also exhibit small variations in nuclides produced by the slow-neutron capture process (*s*-process) of nucleosynthesis, whose pure signature is found in presolar grains originating from asymptotic giant branch (AGB) stars, which are the low-mass to medium-mass stars at the end of their lifetime.

The presence of nucleosynthetic anomalies in materials that formed in the early Solar System implies that presolar grains, or components forming from them, where not homogeneously distributed in the solar accretion disk. This makes nucleosynthetic isotope anomalies in meteorites a unique tracer of transport and mixing processes in the early Solar System and allows genetic relationships among planetary materials to be identified. The impact of nucleosynthetic anomalies on our understanding of early Solar System processes is comparable to what introducing DNA methods meant for forensics: it is a game changer!

## *A Tale of Two Families: The NC–CC Meteorite Isotopic Dichotomy*

Nucleosynthetic anomalies have transformed meteorite classification and our understanding of the dynamical evolution of the asteroid belt. They revealed the existence of an overarching isotopic bimodality among all planetary bodies. This dichotomy is not apparent from the petrography-based classification scheme of meteorites, indicating that the asteroid belt is populated by bodies that formed in spatially distinct locations in the disk, possibly inside and outside of Jupiter's orbit.

The concept and implications of the isotopic dichotomy was first generalized by Warren (2011) who, based on O, Ti, and Cr isotope anomalies, showed that meteorites fall in two distinct populations in multielemental isotope space (FIG. 2). These populations were termed the noncarbonaceous (NC) and carbonaceous (CC) supergroups, because the latter include the carbonaceous chondrites, and the former all the other known chondrite groups. However, both groups also contain various types of volatile-depleted and volatile-rich differentiated meteorites, such that the isotopic grouping is not mirrored in a sample's petrographic or geochemical classification. This reveals the power of nucleosynthetic isotope anomalies as genetic tracers. In contrast to elemental concentrations or mass-dependent isotope variations, they see right through planetary processes (e.g., melting, crystallization, or evaporation), and retain the genetic isotopic fingerprint of their source reservoir.

Since Warren (2011), the isotopic dichotomy has been confirmed and expanded by additional elements (e.g., Ca, Ni, Mo, and Ru). In particular, Mo has provided key insights into the dynamical evolution of the early Solar System. This is because Mo isotopic anomalies are relatively large, and Mo is present in measurable quantities in virtually all meteoritic materials. Furthermore, in $\varepsilon^{95}$Mo–$\varepsilon^{94}$Mo isotope space (FIG. 2B) all measured meteorite data plot on separate, approximately parallel *s*-process mixing lines, the NC and CC lines, so that a single measurement allows for a genetic classification. The Mo data, thus, significantly expanded the number of parent bodies for which an NC–CC classification exists, in particular through the addition of iron meteorites.

The iron meteorites also helped to constrain whether the NC–CC difference is due to a temporal or a spatial heterogeneity in the disk. In other words, whether the NC and CC planetesimals in the asteroid belt formed there in situ, but at different times and with different starting material, or whether they formed in different places in the disk and were only mixed later. Answering this question requires combining the nucleosynthetic anomalies in meteorites with accretion ages of their parent bodies. The Hf–W chronology of NC iron meteorites reveals that their parent bodies started accreting within ~0.5 My of Solar System formation, whereas the parent bodies of CC irons accreted within ~1 My (FIG. 2C). While this finding would allow for some early temporal evolution of disk composition, it is important to recognize that there are chondrites with NC and CC isotopic composition, respectively, whose parent bodies formed between 1 to 3 My after the iron meteorite parent bodies (their later formation is what prevented them from melting, because by then most of the heat-producing, short-lived radionuclide $^{26}$Al had decayed). Taken together, this demonstrates that the formation of NC and CC bodies overlapped for an extended period of time, during which their respective isotopic compositions remained essentially unchanged (FIG. 2C). This indicates that NC and CC bodies formed in contemporaneous but spatially separated disk reservoirs and that at least between ~0.5 My to ~2 My after the start of the

Solar System (the time between the formation of the NC irons and the NC chondrites) there was no significant influx of CC material into the NC reservoir (Kruijer et al. 2017).

How did the NC and CC reservoirs stay separated during ~2 My of planetesimal formation in a dynamic circumsolar disk? How and when were the NC and CC bodies mixed together into the asteroid belt? The answer to these questions is likely the formation of Jupiter, which is not only the largest planet of our Solar System (being some 318 Earth masses), but is also the closest to the asteroid belt. To accrete its gaseous envelope, Jupiter's solid core of 10–20 Earth masses must have formed before dispersal of the gaseous disk. From astronomical observations, we know that the lifetime of gas disks is <10 My, and even 1 My disks already show gaps that may result from the formation of planets (see title figure). As such, it has been proposed that the formation of Jupiter is the great divide: NC bodies form inside its orbit in the gas disk, CC bodies form outside (Warren 2011). Furthermore, the growth and possible migration of Jupiter in the disk also naturally explains the scattering and mixing of NC and CC bodies into the asteroid belt (Raymond and Izidoro 2017). Combined with the age constraints of NC and CC meteorites, this allows a model age for the formation and growth of Jupiter to be constrained (Kruijer et al. 2017). If correct, Jupiter's solid core must have grown to ~20 Earth masses by ~1 My after the start of the Solar System, thereby preventing the influx of CC dust to the NC reservoir before the NC chondrites started to form. This is followed by ~2 My of more protracted growth, until ~50 Earth masses, when Jupiter opened a gap in the disk and efficiently scattered planetesimals around. Jupiter's early formation, thus, most likely shaped and dominated the dynamical and compositional evolution of the inner disk, including the formation and makeup of the terrestrial planets.

## *Isotopic Clues on Terrestrial Planet Formation*

How did the terrestrial planets form? What is the nature and origin of their building blocks? These seemingly simple first-order questions are less well-understood than one might think. It is clear that the planets formed from disk material, and so have a bulk nonvolatile elemental composition similar to the Sun and chondrites. But this is about where the consensus ends. There are fierce debates about which meteorite groups best represent Earth's composition and whether the composition of the accreting material changed with time, as well as debate on the timescale and even the fundamental physical process of terrestrial planet formation itself. For example, the classical model states that the terrestrial planets formed over tens of millions of years—long after the gas disk dissipated—via stochastic collisions among planetary embryos, which themselves formed by runaway growth from local planetesimals (Wetherill 1994). But more recently it has been argued that the terrestrial planets may have formed efficiently by accreting large amounts of inward drifting dust and pebbles: this process requires the presence of gas, and so in the "pebble accretion" model the planets would have formed largely within the ~5 My lifetime of the gaseous disk (Levison et al. 2015). Obviously, these are fundamentally different ways of making the terrestrial planets. Likewise, Earth's isotopic similarity to enstatite chondrites has been used to argue for an isotopically homogeneous accretion of Earth through time (Dauphas 2017), but chemical evolution models for core–mantle differentiation (Rubie et al. 2015), or isotope anomalies in certain elements (Warren 2011; Schiller et al. 2018), have instead been used to argue for heterogeneous accretion, with a change from reduced (inner Solar System) to oxidized (outer Solar System) materials in Earth's accretionary mix through time.

With their ability to genetically trace planetary materials, nucleosynthetic anomalies have the potential to alleviate and eventually resolve these discussions. However, a global model explaining all observations in a coherent way is still lacking. As of now, nucleosynthetic anomalies partly add to the confusion, because different studies use anomaly data to argue for and against certain models. It is, however, only a matter of time until all the puzzle pieces fall into place. The corner-pieces are three key observations. *First*: Different elements inform on different stages of accretion. Because nucleosynthetic isotope anomalies are conserved during melting, the surface rocks of Earth and Mars can be taken as representative of the isotopic composition of the bodies' silicate mantles. Which part of accretion this composition represents depends on the geochemical behavior of the element investigated. For lithophile elements that do not partition into the metallic core (e.g., Ca, Ti, Nd), the mantle isotopic composition represents the full integrated accretion history of the body. But for siderophile elements (e.g., Ni, Fe, Mo, Ru), the mantle isotopic composition is skewed towards the later stages of accretion, because for these elements the earlier-accreted signatures are lost to the core. This means that the rare earth elements (REEs) in your mobile phone are from all of Earth's accretion, the Mo from the last ~15%, and the Pt from the last ~1% of accretion. The concept of combining the geochemical character and nucleosynthetic anomalies of an element to infer the origin of accreting material through time has been elegantly summarized in Dauphas (2017). However, when solving for Earth's accretion history by comparing its mantle composition to those of chondrites this study neglected that *Second*: For some elements, Earth's mantle has an endmember isotopic composition with respect to known meteorites (FIGS. 2, 3). This highlights that Earth must have accreted material not sampled by known meteorites (Burkhardt et al. 2011). Determining the composition of this missing component, and whether it derives from the NC or CC reservoir, will be key for coherently resolving Earth's accretion history. *Third*: Isotopic anomalies among the NC bodies are correlated and Earth's mantle falls on or close to these correlations for all element pairs (FIG. 3). The correlations in the NC reservoir are irrespective of the nucleosynthetic origin of the anomalies, i.e., whether they trace back to *r*-process or *s*-process material, and hold true for all elements, regardless of their geochemical or cosmochemical character. This implies that the isotopic variability in the NC reservoir is governed by simple two-component mixing. Whether this mixing represents a fossilized primordial heliocentric gradient, or a temporal evolution of the inner Solar System, or a mixture of both, remains to be seen. However, the lack of a clear correlation between isotopic anomalies and accretion ages of NC planetesimals seems inconsistent with a temporal change, which supports the heliocentric gradient interpretation (Spitzer et al. 2020). If correct, this would imply that the missing component in Earth represents NC material originating from closer to the Sun than that sampled by known meteorites. In diluted form, some of this missing NC material may be discernible in the small positive Ru isotope anomalies preserved in some Archean rocks (Fischer-Gödde et al. 2020). Finally, even if Earth was largely built from local NC material, the Mo isotope composition of Earth's mantle in-between the NC and CC lines (FIG. 2B) requires that some CC material must have been added to Earth during its later accretion stages (Budde et al. 2019).

In summary, there is currently no consensus model of terrestrial planet formation that coherently explains all the observables. Nevertheless, nucleosynthetic anomalies revealed that Earth contains material not sampled by known meteorites, most likely from the innermost disk, and that it also received some outer Solar System material, at least during the later stages of its growth.

## *Is Everything Preset? Mixing and Disk Building*

To constrain material transport, mixing, and genetic relations in the early Solar System it is important to understand how the isotopic variability among bulk meteorites was first established. This requires a deep dive into the isotopic and chemical compositions of chondrite components and their relation to the isotopic heterogeneity of the Solar System's parental molecular cloud. Isotopically, CC bodies are offset from the NC reservoir towards the composition of CAIs (Figs. 2, 3). Because CAIs carry nucleosynthetic anomalies and are abundant in CC chondrites, but are essentially absent in NC chondrites, this observation is not particularly surprising. However, the CC/NC isotopic offset is seen for all elements showing anomalies, including those not enriched in CAIs. This implies that the CC reservoir is the result of mixing between NC material and a reservoir whose isotopic composition is similar to CAIs, but whose chemical make-up is overall about solar. In this so-called IC (for inclusion-like chondritic) reservoir the CAIs are the refractory high-temperature component. The complementary, less-refractory IC material is less easy to identify, but its presence has been established in individual CC chondrules whose isotopic anomalies were found to vary in-between NC and IC compositions (Schneider et al. 2020; Williams et al. 2020).

As the Solar System's oldest solids, CAIs are thought to have formed at high-temperatures, near the young Sun, but are found in bodies that formed in the outer disk. This requires a process that transported the IC component to the outer Solar System where it was diluted by NC material and produced the CC reservoir. This process may be the formation of the disk itself. Indeed, coupling isotopic anomalies to disk formation models can explain many observables of the meteorite record (Fig. 4). In this simple model, the isotopic difference between the NC and IC reservoirs is a primordial signature inherited from an isotopically heterogeneous molecular cloud. During the initial stages of cloud collapse, the infalling material is limited to regions close to the forming Sun. This early infalling material must have been characterized by an IC composition and experienced high-temperature processing close to the Sun before being transported outwards by viscous spreading of the forming disk. With time, the infalling material changed to an NC isotopic composition, diluting the IC signature of the early disk, particularly in the inner disk regions where most of the infalling mass is deposited. However, with expanding centrifugal radius, later infall also added unprocessed NC cloud material to the outer disk. At the end of the infall-stage, the inner part of the disk contained higher amounts of thermally processed NC materials while the outer disk contained a higher proportion of processed IC materials and also a higher proportion of unprocessed primitive dust and an isotopic composition between the NC and IC reservoirs. These early set gradients in the disk were then retained throughout the accretion of planetesimals in the NC and CC reservoirs, likely aided by the early formation of Jupiter. Thus, to a first order, the isotopic variability between NC and CC bodies, along with many of their petrologic and compositional properties, can be explained by simple mixing of two primordial disk reservoirs. The selective processing of these reservoirs in different nebular environments, and the heterogeneous distribution of the resulting nebular products, then led to the observed isotopic and elemental variations within the NC and CC reservoirs. As such, much of the compositional diversity of meteorites has its origin in the very initial stages of disk building and evolution (Burkhardt et al. 2019; Nanne et al. 2019; Jacquet et al. 2019).

## Correlations Among Mass-Dependent and Mass-Independent Isotope Effects?

Whereas nucleosynthetic isotope anomalies trace Solar System source reservoirs and mixing, mass-dependent isotope fractionation can be used to investigate planetary or nebular processes such as condensation, evaporation, or melting. Given the fundamentally different nature of mass-dependent isotope variations and nucleosynthetic anomalies, correlations between them are not necessarily expected. Yet, such correlations exist, in particular with mass-dependent isotope variations in moderately volatile elements like Zn, Rb, or Te (e.g., Luck et al. 2005; Pringle et al. 2017; Hellmann et al. 2020). While keeping in mind that correlation does not imply causality, these relationships nevertheless offer a fresh look at the interpretation of element concentration or mass-dependent isotope data among meteorites and their components. For example, the origin of systematic volatile element depletions among CC chondrites is a matter of long-standing debate and has been attributed to either incomplete condensation in the solar nebula or variable mixing between volatile-rich and volatile-poor components. The correlation between volatile element concentration, isotope fractionation, and nucleosynthetic anomalies shown in FIGURE 5 for $\delta^{128/126}$Te and $\epsilon^{54}$Cr strongly argues for the mixing scenario. Mass-dependent isotope variations in moderately volatile elements among CC chondrites are, therefore, not a direct feature of condensation but the result of variable physical mixing between isotopically light, volatile-poor chondrules (or their precursors), and isotopically heavy, volatile-rich matrix from different nebular source regions.

While the systematic search for correlations between mass-dependent and mass-independent isotope variations among planetary materials is just beginning, it ultimately has the potential to delaminate signatures of nebular physical mixing from those of planetary processes. Distinguishing these two is critical for the correct interpretation of mass-dependent isotope variations among planetary materials, in particular for elements with just two isotopes (e.g., N, K, V, Cu, Ga, Rb, Ag, or Sb).

# CONCLUSIONS AND OUTLOOK

Isotope anomalies are currently revolutionizing our understanding of the Solar System's dynamical evolution, from the collapse of the molecular cloud to the formation of the terrestrial planets. As genetic tracers, these anomalies reveal an overarching isotopic bimodality among planetary bodies that ultimately dates back to the infall stage of Solar System formation, allow to place a model age on the formation of Jupiter, demonstrate that Earth accreted material from an unsampled reservoir, and by differentiating between process and source signatures can aid the interpretation of mass-dependent isotope data. Many open questions remain, but, clearly, planetary isotope genetics has the potential to bring meteoritics, astronomical observations, and modeling together towards a unified model for early Solar System evolution.

# ACKNOWLEDGMENTS

I thank the guest editors for their invitation, the cosmochemistry group at the IfP for discussions, and H. Palme and an anonymous reviewer for helpful comments. Funded by the Deutsche Forschungsgemeinschaft (DFG, German Research Foundation)–Project-ID 263649064–TRR170. This is TRR170 pub. no. 151.

# FIGURES

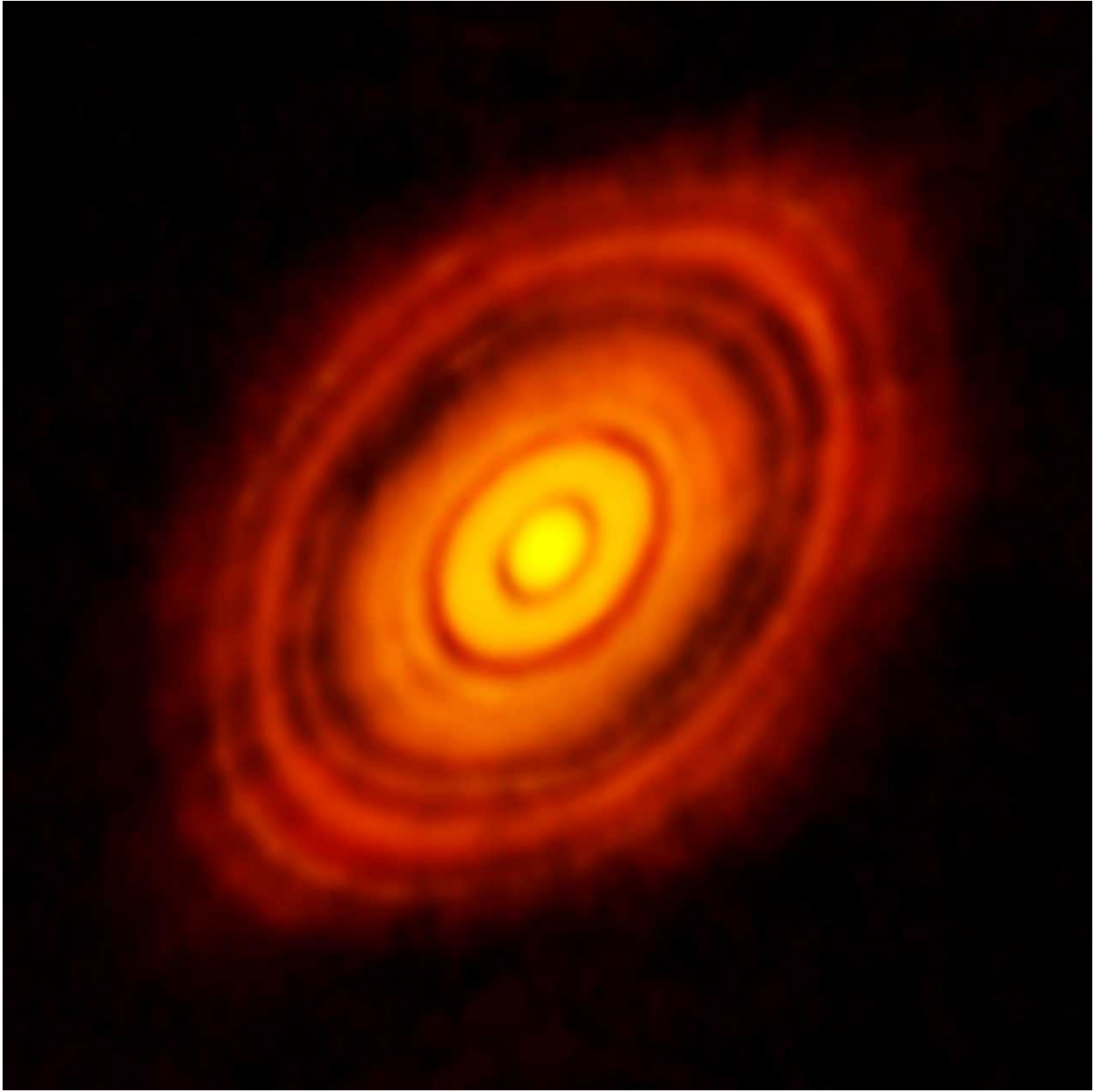

**Title Figure** Image of the less than one-million year old star HL Tauri and its accretion disk. Dark concentric rings indicate planet formation. CREDIT: ALMA (ESO/NAOJ/NRAO).

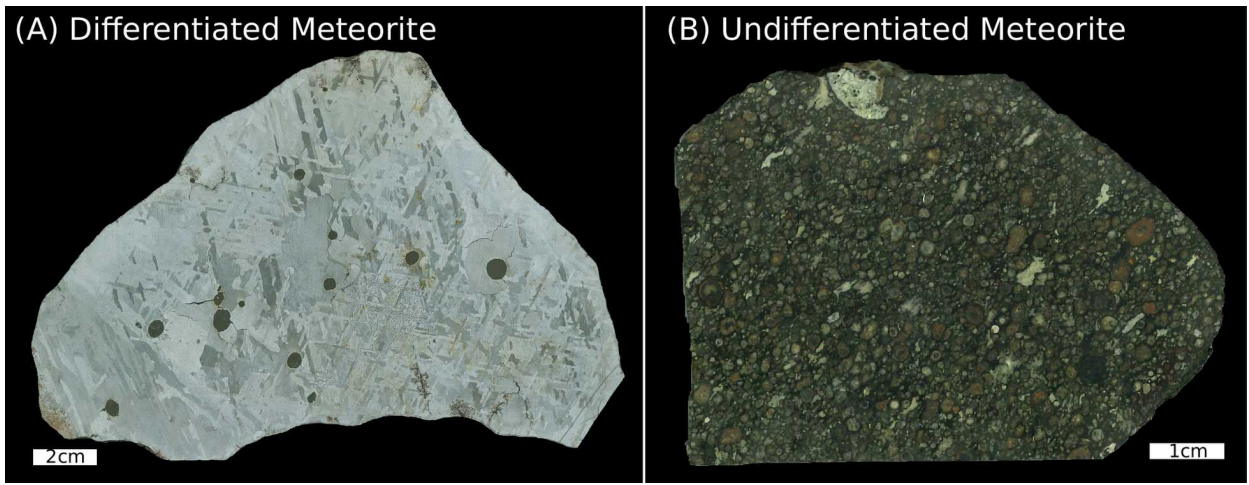

**FIGURE 1** Meteorites are invaluable samples to constrain the origin and evolution of the Solar System. (**A**) Differentiated meteorites sample the crust, mantle, or core material from differentiated planetesimals. Pictured here is an example of core material (the Mt. Dooling meteorite). (**B**) Undifferentiated meteorites, such as the chondrite NWA 8333, sample undifferentiated bodies in which nebular materials such as chondrules, Ca,Al-rich inclusions (CAIs), and matrix have survived since their accretion.

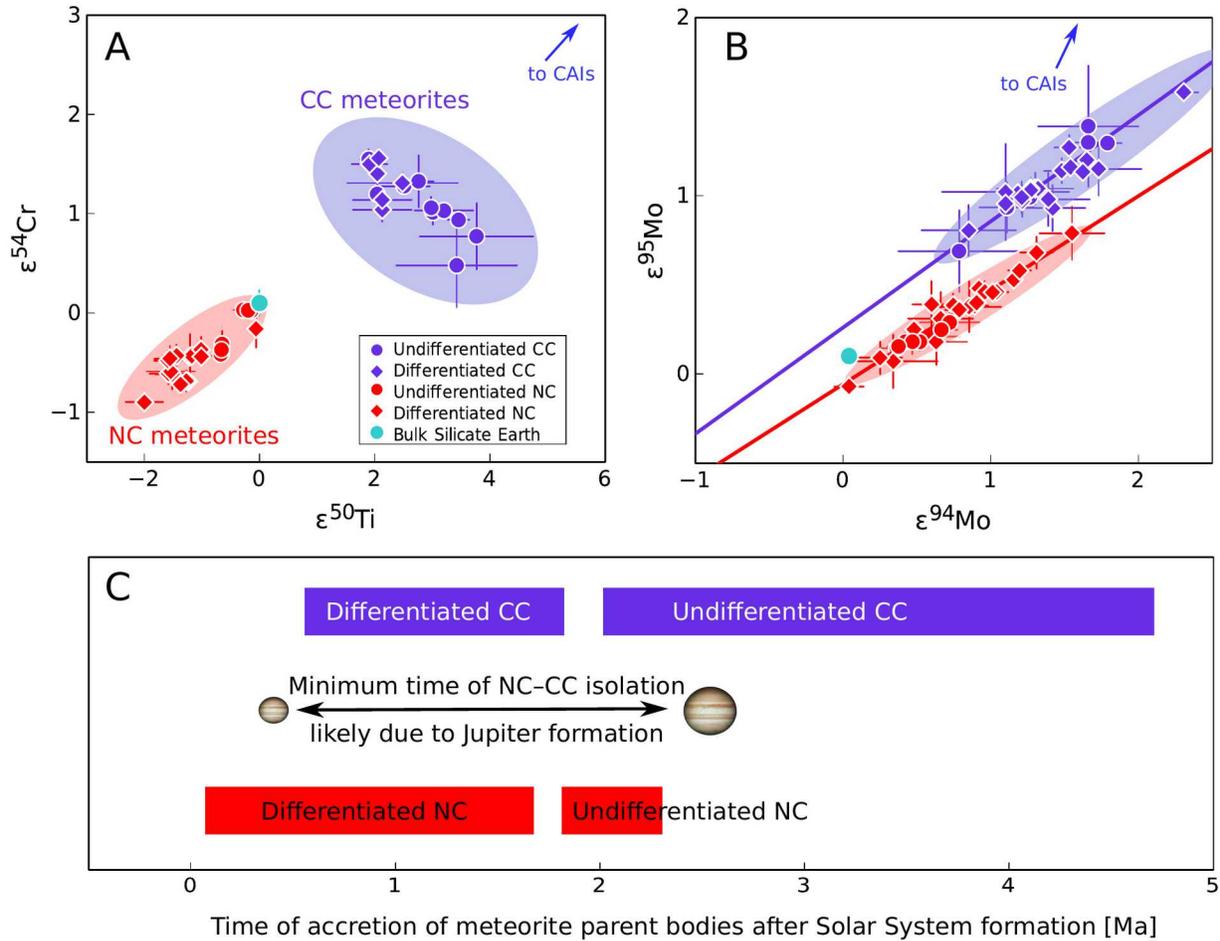

**FIGURE 2** Nucleosynthetic anomalies of meteorites reveal a fundamental dichotomy among planetary bodies in the early Solar System. (**A**) In an $\varepsilon^{50}$Ti vs. $\varepsilon^{54}$Cr plot, meteorites define two separate fields: the NC (noncarbonaceous) and CC (carbonaceous) meteorite fields (see Ibañez-Mejia and Tissot 2021 this issue for $\varepsilon$-notation). CAI = Ca,Al-rich inclusion. (**B**) In $\varepsilon^{95}$Mo vs. $\varepsilon^{94}$Mo space, the NC and CC meteorites define two near-parallel lines. In both diagrams, CC meteorites are offset from NC meteorites towards the composition of CAIs. (**C**) When coupled with age constraints, the dichotomy reveals that spatially separated, isotopically distinct reservoirs coexisted for a prolonged time in the accretion disk around the young Sun.

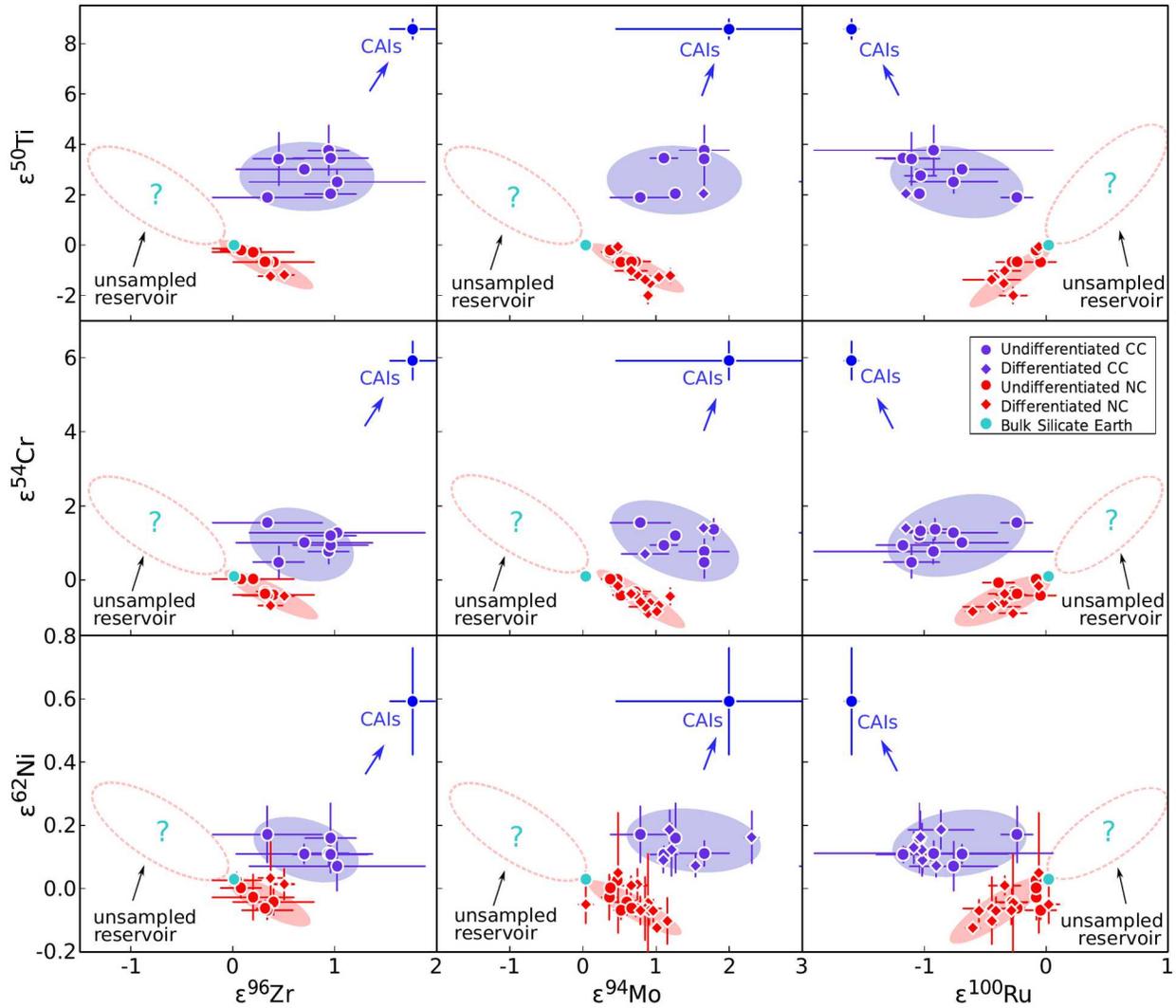

**FIGURE 3** The composition of Earth's mantle (bulk silicate Earth, BSE) in relation to the isotope anomalies of meteorites in multielemental isotope space. The position of the BSE at the end of the NC (noncarbonaceous meteorite) correlations reveals that Earth must have accreted material not sampled by known meteorites, most likely from the NC reservoir sunwards of Earth's orbit. CAIs = Ca,Al-rich inclusions; CC = carbonaceous meteorites.

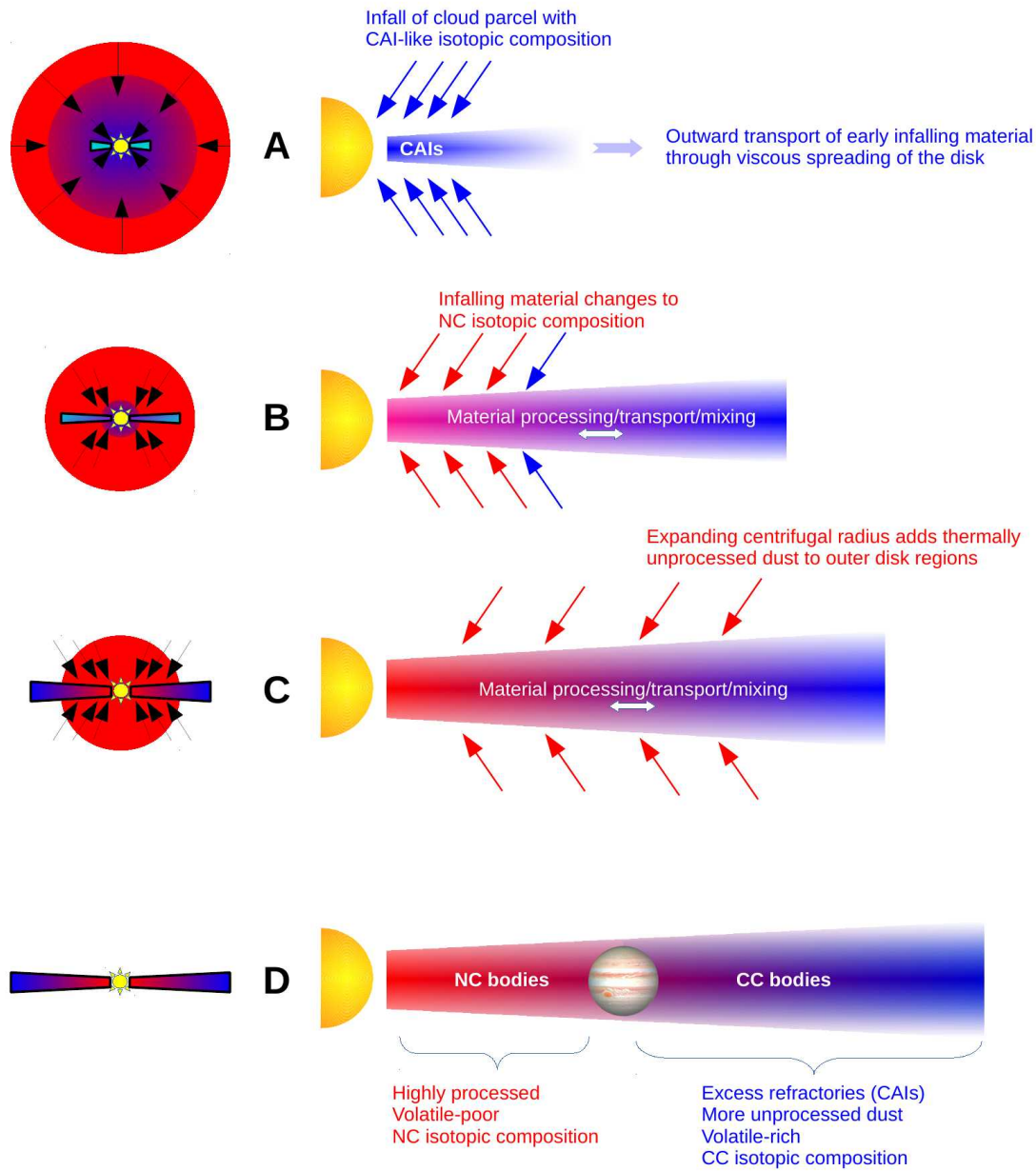

**FIGURE 4** The origin of the noncarbonaceous–carbonaceous meteorite (NC–CC) dichotomy probably dates back to the disk-building phase of the Solar System. (**A**) Very early phase where early infalling material had an isotopic composition similar to Ca,Al-rich inclusions (CAIs). Through viscous spreading of the disk, this so-called IC (inclusion-like chondritic) material was transported to the outer regions of the disk. (**B**) Later infalling material dominates the inner disk regions and has an NC isotopic composition. (**C**) The CC composition results from mixing between the NC and IC reservoirs. (**D**) The early formation of Jupiter prevents further exchange so that bodies forming inside of Jupiter retained an NC composition, and the ones forming outside of Jupiter a CC composition. CREATIVE COMMONS BY BURKHARDT ET AL. (2019).

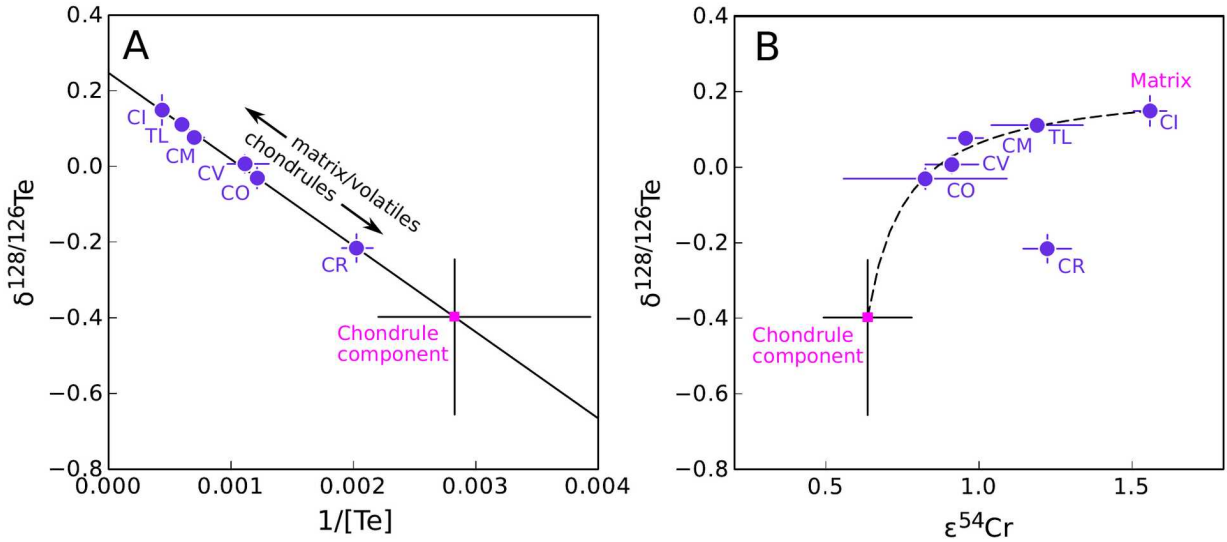

**FIGURE 5** The origin of volatile element depletion trends among carbonaceous chondrites is debated. The correlations between elemental depletion and mass-dependent isotope variations (shown in **A**), and nucleosynthetic anomalies (shown in **B**) implies that the volatile depletion trends are not due to nebular fractionation but are due to mixing (dashed line) between isotopically light, volatile-poor chondrules (or their precursors) and isotopically heavy, volatile-rich matrix from different nebular source regions. The chondrule component of CR chondrites is likely sourced from material with higher $\varepsilon^{54}$Cr. Abbreviations: CI = carbonaceous chondrite of the Ivuna type; TL = carbonaceous chondrite Tagish Lake; CM = carbonaceous chondrite of the Mighei type; CV = carbonaceous chondrite of the Vigarano type; CO = carbonaceous chondrite of the Ornans type; CR = carbonaceous chondrite of the Renazzo type. MODIFIED AFTER HELLMANN ET AL. (2020).